\documentclass[apjl]{emulateapj}

\usepackage{graphicx}
\usepackage{t1enc}
\usepackage[varg]{txfonts}
\usepackage{epsfig}
\usepackage{rotating}
\usepackage{natbib}
\usepackage{color}

\newcommand{\mum}   {$\mu$m}
\newcommand{\kms}   {km~s$^{-1}$}

\newcommand{\jpb}   {$\rm Jy~beam^{-1}$}    
\newcommand{\lo}    {$L_{\sun}$}
\newcommand{\mo}    {$M_{\sun}$}

\newcommand{\methanol}  {CH$_3$OH}
\newcommand{\water}  {H$_2$O}
\newcommand{\otcs} {O$^{13}$CS}
\newcommand{\ethyleneg}  {(CH$_2$OH)$_2$}
\newcommand{\ethanol}  {CH$_3$CH$_2$OH}
\newcommand{\acetone} {CH$_3$COCH$_3$}
\newcommand{\dimethylether} {CH$_3$OCH$_3$}
\newcommand{\acetaldehyde} {CH$_3$CHO}
\newcommand{\methylformate} {CH$_3$OCHO}

\newcommand{\methylcyanide} {CH$_3$CN}
\newcommand{\vinylcyanide} {CH$_2$CHCN}
\newcommand{\ethylcyanide} {CH$_3$CH$_2$CN}
\newcommand{\et}    {et al.}
\newcommand{\eg}    {e.\,g.,}
\newcommand{\ie}    {i.\,e.,}

\newcommand{\phe}   {\phantom{$^\mathrm{c}$}}

\definecolor{RED}{rgb}{1.0,0.0,0.0}

\shorttitle{Intermediate-mass hot cores at $\sim500$~AU}
\shortauthors{Palau et al.}

\begin{document}

\title{Intermediate-mass hot cores at $\sim500$~AU: disks or outflows?\footnote{Based on observations carried out with the IRAM Plateau de Bure Interferometer. IRAM is supported by INSU/CNRS (France), MPG (Germany) and IGN (Spain).}
}


\author{Aina Palau\altaffilmark{2},
Asunci\'on Fuente\altaffilmark{3},
Josep M. Girart\altaffilmark{2},
Francesco Fontani\altaffilmark{4},
J\'er\'emie Boissier\altaffilmark{5,6},
Vincent Pi\'etu\altaffilmark{7},
\'Alvaro S\'anchez-Monge\altaffilmark{4},
Gemma Busquet\altaffilmark{8},
Robert Estalella\altaffilmark{9},
Luis A. Zapata\altaffilmark{10},
Qizhou Zhang\altaffilmark{11},
Roberto Neri\altaffilmark{7},
Paul T. P. Ho\altaffilmark{11,12},
Tom\'as Alonso-Albi\altaffilmark{3},
Marc Audard\altaffilmark{13}
}
\altaffiltext{2}{Institut de Ci\`encies de l'Espai (CSIC-IEEC), Campus UAB -- Facultat de Ci\`encies, Torre C5 -- parell 2, 08193 Bellaterra, Catalunya, Spain}
\email{palau@ieec.uab.es}
\altaffiltext{3}{Observatorio Astron\'omico Nacional, P.O. Box 112, 28803 Alcal\'a de Henares, Madrid, Spain}
\altaffiltext{4}{Osservatorio Astrofisico di Arcetri, INAF, Lago E. Fermi 5, 50125, Firenze, Italy}
\altaffiltext{5}{Istituto di Radioastronomia, INAF, Via Gobetti 101, Bologna, Italy}
\altaffiltext{6}{ESO, Karl Schwarzschild St. 2, 85748 Garching Muenchen, Germany}
\altaffiltext{7}{IRAM, 300 Rue de la piscine, 38406 Saint Martin d'H\`eres, France}
\altaffiltext{8}{Istituto di Fisica dello Spazio Interplanetario, INAF, Area di Recerca di Tor Vergata, Via Fosso Cavaliere 100, 00133 Roma, Italy}
\altaffiltext{9}{Departament d'Astronomia i Meteorologia (IEEC-UB), Institut Ci\`encies Cosmos, Universitat Barcelona, Mart\'i Franqu\`es, 1, 08028 Barcelona, Spain}
\altaffiltext{10}{Centro de Radioastronom\'ia y Astrof\'isica, Universidad Nacional Aut\'onoma de M\'exico, P.O. Box 3-72, 58090, Morelia, Michoac\'an, Mexico}
\altaffiltext{11}{Harvard-Smithsonian Center for Astrophysics, 60 Garden Street, Cambridge, MA 02138, USA}
\altaffiltext{12}{Institute of Astronomy and Astrophysics, Academia Sinica, P.O. Box 23-141, Taipei 106, Taiwan}
\altaffiltext{13}{Geneva Observatory, University of Geneva, Ch. des Maillettes 51, 1290 Versoix, Switzerland}

\begin{abstract}
Observations with the Plateau de Bure Interferometer in the most extended configuration toward two intermediate-mass star-forming regions, IRAS\,22198+6336 and AFGL\,5142, reveal the presence of several complex organic molecules at $\sim500$~AU scales, confirming the presence of hot cores in both regions. The hot cores are not rich in CN-bearing molecules, as often seen in massive hot cores, and are mainly traced by \ethanol, \ethyleneg, \acetone, and \methanol, with additionally \acetaldehyde, CH$_3$OD and HCOOD for IRAS\,22198+6336, and C$_6$H,  and \otcs\ for AFGL\,5142. The emission of complex molecules is resolved down to sizes of $\sim300$ and $\sim600$~AU, for IRAS\,22198+6336 and AFGL\,5142, respectively, and most likely is tracing protostellar disks rather than flattened envelopes or toroids as usually found. This is specially clear for the case of IRAS\,22198+6336, where we detect a velocity gradient for all the mapped molecules perpendicular to the most chemically rich outflow of the region, yielding a dynamic mass  $\gtrsim4$~\mo. As for AFGL\,5142, the hot core emission is resolved into two elongated cores separated $\sim1800$~AU.
A detailed comparison of the complex molecule peaks to the new CO\,(2--1) data and \water\ maser data from literature suggests that also for AFGL\,5142 the complex molecules are mainly associated with disks, except for a faint and extended molecular emission found to the west, which is possibly produced in the interface between one of the outflows and the dense surrounding gas. 
\end{abstract}


\section{Introduction \label{sint}}

\begin{figure*}[t!]
\begin{center}
\begin{tabular}[b]{cc}
    \epsfig{file=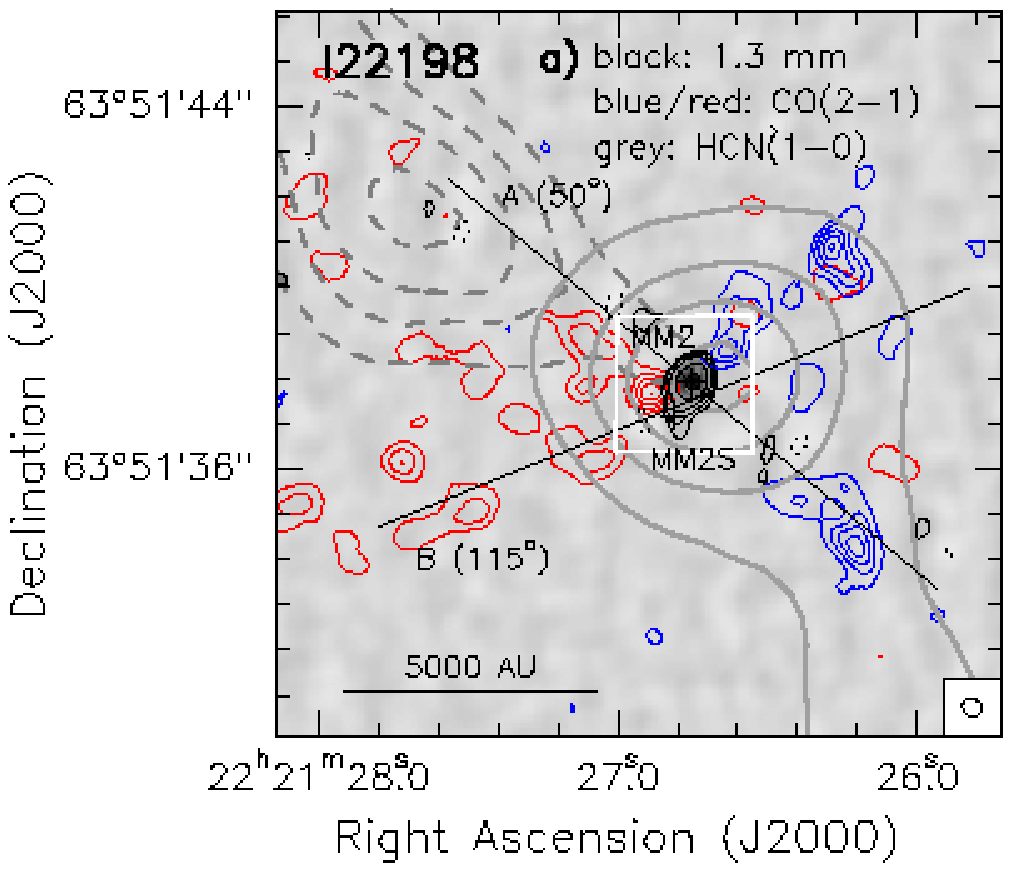, scale=0.75}&
    \epsfig{file=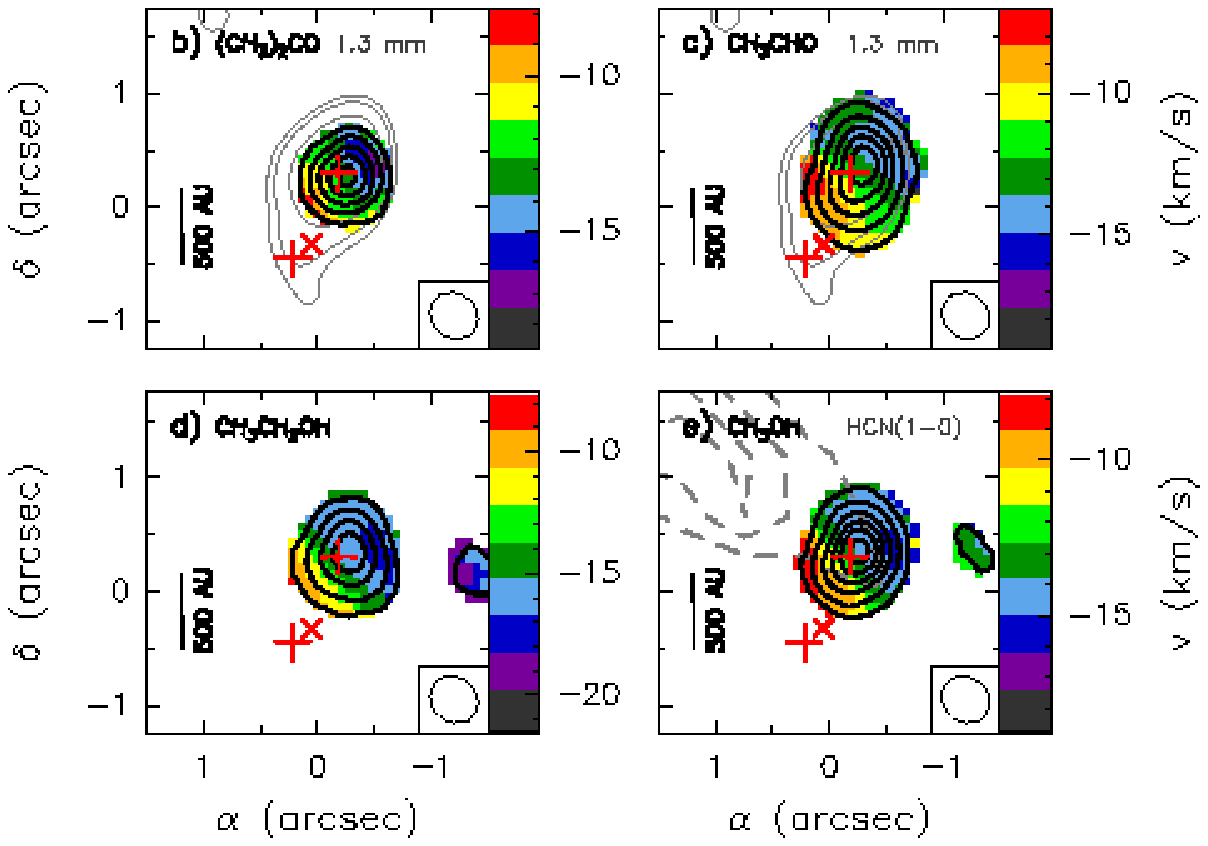, scale=0.75}\\
\end{tabular}
\caption{I22198 region.
 {\bf a)} Greyscale and black contours are the 1.3~mm continuum emission (natural weight; beam: $0.43''\times0.39''$, P.A.=$27.1^\circ$; rms: 2.0~m\jpb). Contour levels are 4, 6, 10, 20, and 40 times 2~m\jpb. Grey contours are the red(dashed) and blue(solid)-shifted emission of HCN\,(1--0) tracing outflow A (S\'anchez-Monge 2011).
Blue/red contours: CO\,(2--1) emission integrated from $-31.0$ to $-19.7$/$-5.3$ to 7.9~\kms\ (beam: $0.55''\times0.50''$, P.A. =$42.5^\circ$, tapered to 400~m).
Blue/red contour levels start at 20\%/18\% of the maximum, 1754.97/1766.58~\jpb\ m\,s$^{-1}$,  and increase in steps of 20\%.
{\bf b--e)} Black thick contours/colorscale: zero/first-order moment of different COMs (at 230.176 (b),  230.316 (c), 230.231 (d), and 230.368~GHz (e); beam: $0.44''\times0.39''$, P.A. =$22.3^\circ$). Contour levels are 3, 10, 20, 30, 40, 50, and 60 times 15~\jpb\ m\,s$^{-1}$.
Black thin contours:  {\bf b-c)} 1.3~mm continuum emission. {\bf e)} red-shifted HCN\,(1--0) emission (same as panel a). 
In all panels, red plus signs correspond to the millimeter continuum peaks, and the cross to the Spitzer-IRAC (3.6~$\mu$m) source in the field (S\'anchez-Monge, priv. commun.).
}
\label{flit3}
\end{center}
\end{figure*}


Hot molecular cores are compact ($\le\!0.05$~pc) objects with high temperatures ($\gtrsim100$~K) and densities ($n\gtrsim\!10^6$~cm$^{-3}$), 
which are characterized by a very rich chemistry of complex organic molecules (COMs, molecules with more than 6 atoms) 
and are typically associated with deeply embedded massive protostars (\eg\ Cesaroni 2005).
Such a rich chemistry is supposed to be triggered by the radiation from the nascent massive star which evaporates the complex molecules from the dust grain mantles and triggers additional gas-phase reactions (\eg\  Millar \et\ 1997). Thus, hot cores are supposed to be radiatively heated and originate in the innermost parts of the condensation where the massive star is being formed.
However, the radiative heating mechanism is questioned, as some observations suggest that the COM emission could be associated with shocks as well (Liu \et\ 2002; Chandler \et\ 2005; Goicoechea \et\ 2006;  Goddi \et\ 2011a; Favre \et\ 2011a; Zapata \et\ 2011).
This question cannot be easily answered because most of the hot cores known to date ($\sim100$) are associated with high-mass protostars, and are on average located at distances $> 2$~kpc (studied typically at spatial scales $\gtrsim4000$~AU), while in the low-mass case (studied at scales 500--1000~AU, \eg\ Bisschop \et\ 2008), there are very few clear cases (the so-called hot corinos: \eg\ Ceccarelli 2004; Bottinelli \et\ 2007). 
Concerning the intermediate-mass regime, the only well-known cases are NGC\,7129 (Fuente \et\ 2005), IC\,1396N (Fuente \et\ 2009), 
and IRAS\,22198+6336 (S\'anchez-Monge \et\ 2010), and still the presence of COMs is not clear in all the cases, questioning the true nature of these objects as hot cores. Thus, a detailed high spatial resolution ($\sim500$~AU) study of true hot cores (\ie\ with complex organic chemistry) is lacking.


In this letter, we present new subarcsecond interferometric observations of COMs towards two intermediate-mass star-forming regions containing hot core candidates.
The two regions are IRAS~22198+6336 (hereafter I22198), an intermediate-mass protostar of 370~\lo\ located at 760~pc  
(Hirota \et\ 2008), deeply embedded and driving a quadrupolar outflow with a thermal radiojet at its base (S\'anchez-Monge \et\ 2010), and AFGL\,5142 (hereafter A5142), a protocluster of 2300~\lo\ (S\'anchez-Monge 2011) located at 1.8~kpc, driving at least 3 outflows and associated also with thermal free-free emission and with no infrared emission (Hunter \et\ 1995). Submillimter Array observations of \methylcyanide\ at  $\sim2''$ angular resolution toward both regions indicate that the protostars are embedded in dense gas at around 100~K, but no clear evidence of COMs has been reported so far (Zhang \et\  2007; S\'anchez-Monge \et\ 2010). Here we show multiple detections of COMs down to angular scales of $0.4''$.

\begin{figure*}[t!]
\begin{center}
\begin{tabular}[b]{cc}
     \epsfig{file=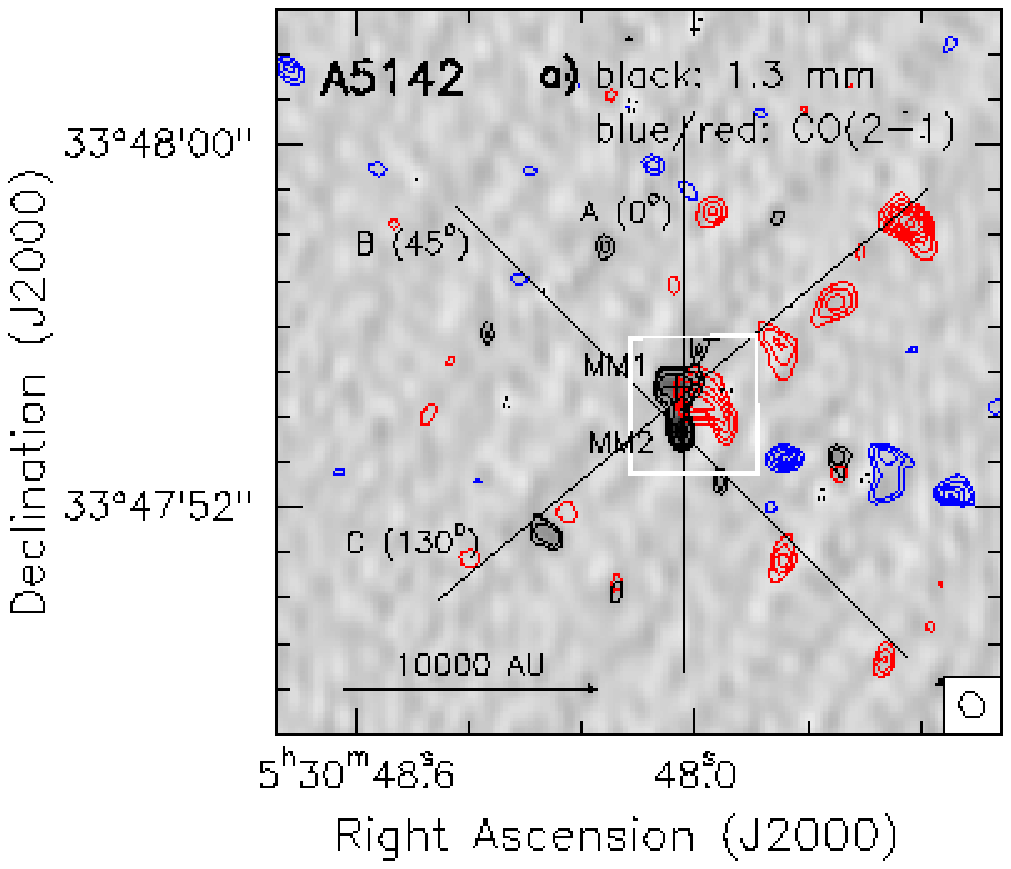, scale=0.75}&
     \epsfig{file=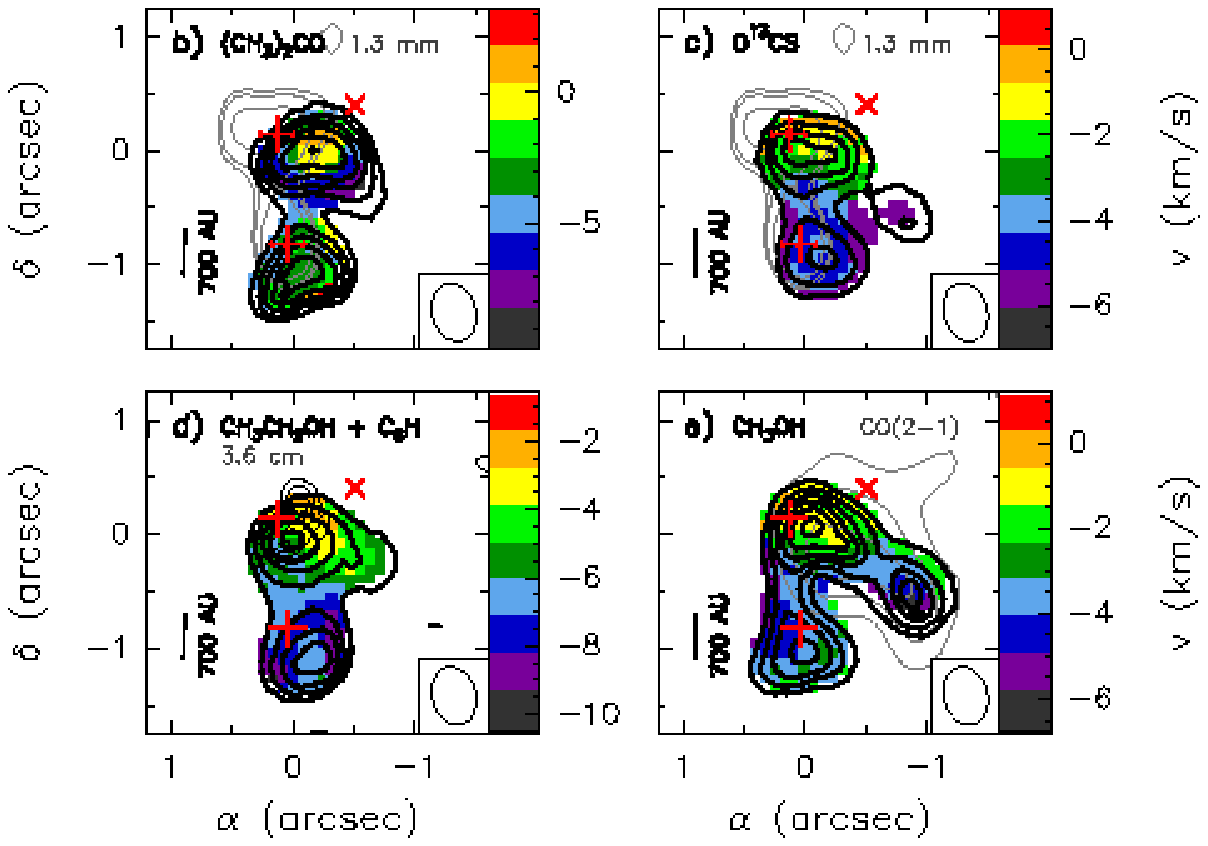, scale=0.75}\\
\end{tabular}
\caption{
A5142 region. 
{\bf a)} Greyscale and black contours are the 1.3~mm continuum emission (uniform weight; beam: $0.50''\times0.30''$, P.A.=2.1$^\circ$; rms: 2.8~m\jpb). Contour levels are 4, 6, 10, 20, and 40 times 2.4~m\jpb.
Blue/red contours: CO\,(2--1) emission integrated from $-19.7$ to $-8.9$/0.7 to 12.7~\kms\ (beam: $0.59''\times0.51''$, P.A. =$27.7^\circ$, tapered to 300~m).
Blue/red contour levels start at 50\%/37\% of the maximum, 1851.04/3175.57~\jpb\ m\,s$^{-1}$,  and increase in steps of 10\%.
{\bf b--e)} Black thick contours/colorscale: zero/first-order moment of different COMs (see Figure~1) (beam: $0.52''\times0.36''$, P.A. =$12.0^\circ$). Contour levels are 4, 8, 12, 16, 20, and 24 times 27 \jpb\ m\,s$^{-1}$.
Black thin contours: {\bf b--c)} 1.3~mm continuum emission; {\bf d)} centimeter continuum (Zhang \et\ 2007); {\bf e)} CO\,(2--1) redshifted emission. 
Symbols as in Figure 1 (Spitzer source from Qiu \et\ 2008).
}
\label{flit3}
\end{center}
\end{figure*}

\section{Observations \label{sobs}}

The IRAM Plateau de Bure Interferometer (PdBI)
was used in its most extended configuration to observe the continuum and CO\,(2--1) emission at 230.538~GHz of I22198 and A5142, 
providing baselines in the range 136--760~m.
A5142 was observed on 2010 January 10th (phase center: 05:30:48.02  +33:47:54.5), with system temperature of 170~K and averaged atmospheric precipitable water vapor around 2~mm. Bandpass calibration was carried out using B0851+202. To calibrate the phases we used B0552+398, and J0512+294, yielding a phase rms of $20^\circ$--$60^\circ$. 
A subsequent track was carried out on 2010 January 18th to observe  in track-sharing mode the I22198 region
(phase center: 22:21:26.78 +63:51:37.6).
The system temperature was $\sim150$~K and the precipitable water vapor was $\sim\,1$~mm. 3C454.3 was used for the bandpass calibration. Phases were calibrated with B2146+608, and B2037+511, and we obtained a phase rms of $20^\circ$--$40^\circ$. 
The absolute flux density scale was determined from 3C273 and B2146+608, with an estimated uncertainty around 30\%.
The estimated uncertainty in absolute position is $0.03''$--$0.05''$.
%
%

A correlator unit  of 40 MHz bandwidth with 256 spectral channels was used to observe the CO\,(2--1) line in each polarization. 
Three additional units (per polarization) of 320~MHz bandwidth with 128 channels were used to observe the continuum across $\sim1$~GHz. 
The COMs were detected in the 320~MHz units, which provide a spectral resolution of 2.5 MHz or 3.25~\kms. 
%
Calibration and imaging were performed using the \textsc{GILDAS} software package\footnote{GILDAS: Grenoble Image and Line Data
Analysis System, see http://www.iram.fr/IRAMFR/GILDAS.}, following the standard procedures.
The  synthesized beams are $0.43''\times0.39''$ at P.A.=$27.1^\circ$ for I22198, and $0.50''\times0.30''$ at P.A.=$2.1^\circ$ for A5142.
The rms of the final cleaned maps are given in the captions of Figures~1 and 2. 
%

\section{Results \label{sres}}


\subsection{Continuum and CO\,(2--1) \label{srescontco}}

In Figures~1-a and 2-a we present the 1.3~mm continuum and CO\,(2--1) emission of I22198 and A5142 down to $0.4''$. 
For I22198 we detected one strong source, MM2 (following the nomenclature in S\'anchez-Monge \et\ 2010), extended in the southeast-northwest direction, with a faint extension at 6$\sigma$, MM2-S, $0.8''$ to the southeast.  The overall extended emission of MM2 is perpendicular to the direction of outflow A (S\'anchez-Monge \et\ 2010). We fitted in the $uv$-plane an elliptical Gaussian to MM2 and obtained residual emission at the position of MM2-S, indicating an additional point source.  The coordinates determined for MM2-S are (J2000): 22:21:26.807, 63:51:37.14, and the flux density is $18.2\pm0.8$~mJy, which corresponds to a mass of 0.1--0.6~\mo, assuming a dust temperature  of 10--30~K, 
a gas-to-dust mass ratio of 100, and a dust mass opacity coefficient at 1.3~mm of 0.899~cm$^2$\,g$^{-1}$
(agglomerated grains with thin ice mantles for densities $\sim10^6$~cm$^{-3}$, Ossenkopf \& Henning 1994). The uncertainty in the masses is estimated to be a factor of 2.
As for I22198-MM2, the deconvolved size  is $500\times300$~AU at P.A.$=-35^\circ$, the peak intensity and flux density are $91.8\pm0.9$~m\jpb, and  $246\pm3$~mJy, and the mass is $\sim1$~\mo\ (Table~1).
%
Towards A5142 the millimeter emission is dominated by two partially extended and strong sources, MM1 and MM2, which are surrounded by five faint point-like sources (Palau \et, in prep.). 
The deconvolved sizes (from elliptical Gaussian fits in the $uv$-plane) are $1200\times900$~AU at P.A.$=-86^\circ$ for MM1 and $1000\times400$~AU at P.A.$=+18^\circ$ for MM2.
The peak intensities and flux densities are $38\pm3$~m\jpb, $212\pm7$~mJy for MM1, and $62\pm3$~m\jpb,  $151\pm4$~mJy for MM2, yielding masses of $\sim4$~\mo\ (Table~1).

%

Regarding the CO\,(2--1) emission, we first caution that an important part of the emission is filtered out by the interferometer and we are only sensitive to compact knots, even after tapering the data to a final beam of $0.6''$. In I22198 we detected chains of knots possibly tracing the cavity walls of outflows A and B (S\'anchez-Monge \et\ 2010). 
As for A5142, the CO\,(2--1) emission is again very clumpy but showing chains of knots which match well the known outflows of the region (\eg\ Zhang \et\ 2007). 


\subsection{Hot molecular core \label{sreshc}}


In Figure~3-left we show the PdBI spectrum towards I22198-MM2, A5142-MM1 and A5142-MM2 for the entire observed bandwidth. Line identification was performed following the methodology described below. First, we searched for molecules with $\geq5$ transitions detectable within the observed frequency range and compared preliminar synthetic spectra (see below) for these molecules to the observed spectra. We found that, for the three sources, \ethanol\ and \ethyleneg\ alone could account for about half of the detected transitions (Fig.~3-left). Second, we computed rotational diagrams (Fig.~4) with the \ethanol\ transitions, allowing us to determine the gas temperature for each region (120~K for I22198, 210~K for A5142-MM1, and 140~K for A5142-MM2). Third, we adopted the derived temperatures, as well as the average linewidths of the involved transitions (given in Fig.~3), to compute synthetic spectra for both \ethanol\ and \ethyleneg, summed them, and subtract the sum to the observed spectrum. Finally, the definitive line identification of molecules different from \ethanol\ and \ethyleneg\ was performed in the residual spectrum (Fig.~3-right) for lines with flux above 4$\sigma$ ($\sigma=4.5$~mJy for I22198; $\sigma=7.5$~mJy for A5142). 
The systemic velocity used for I22198 was derived from the strongest isolated lines and was found to be $-12.3$~kms. As for A5142, we adopted the systemic velocities derived by Zhang \et\ (2007): $-1.0$~\kms\ for A5142-MM1, and $-3.4$~\kms\ for A5142-MM2. 
The synthetic spectra were computed using the estimates of temperature and linewidth given above, and assuming local thermodynamic equilibrium, optically thin emission, and the molecular data from the Jet Propulsion Laboratory (Pickett \et\ 1998) or the Cologne Database for Molecular Spectroscopy catalogs (M\"uller \et\ 2005), except for CH$_3$OD (Anderson \et\ 1988).
The final identification is presented in Fig~3-right and the column densities used to build the synthetic spectra are listed in Table~1. 

In addition to \ethanol\ and \ethyleneg, the strongest lines found in the spectra of the three sources are from \methanol\ and \acetone, and for none of the sources CN-bearing species such as \vinylcyanide,  \ethylcyanide\ were required to fit the spectra. Molecules detected only in I22198 are  \acetaldehyde, \methylformate, CH$_3$OD and HCOOD.
On the other hand, for A5142-MM1, no deuterated species were detected, \otcs\ was found to dominate the line at 230.317~GHz (with no need of \acetaldehyde), and we identfied
two transitions of C$_6$H, with energies of the upper state of 483~K.
The spectrum observed in A5142-MM2 is essentially the same as that of A5142-MM1, with smaller fluxes.

\begin{figure*}[t!]
\begin{center}
\begin{tabular}[b]{c}
    \epsfig{file=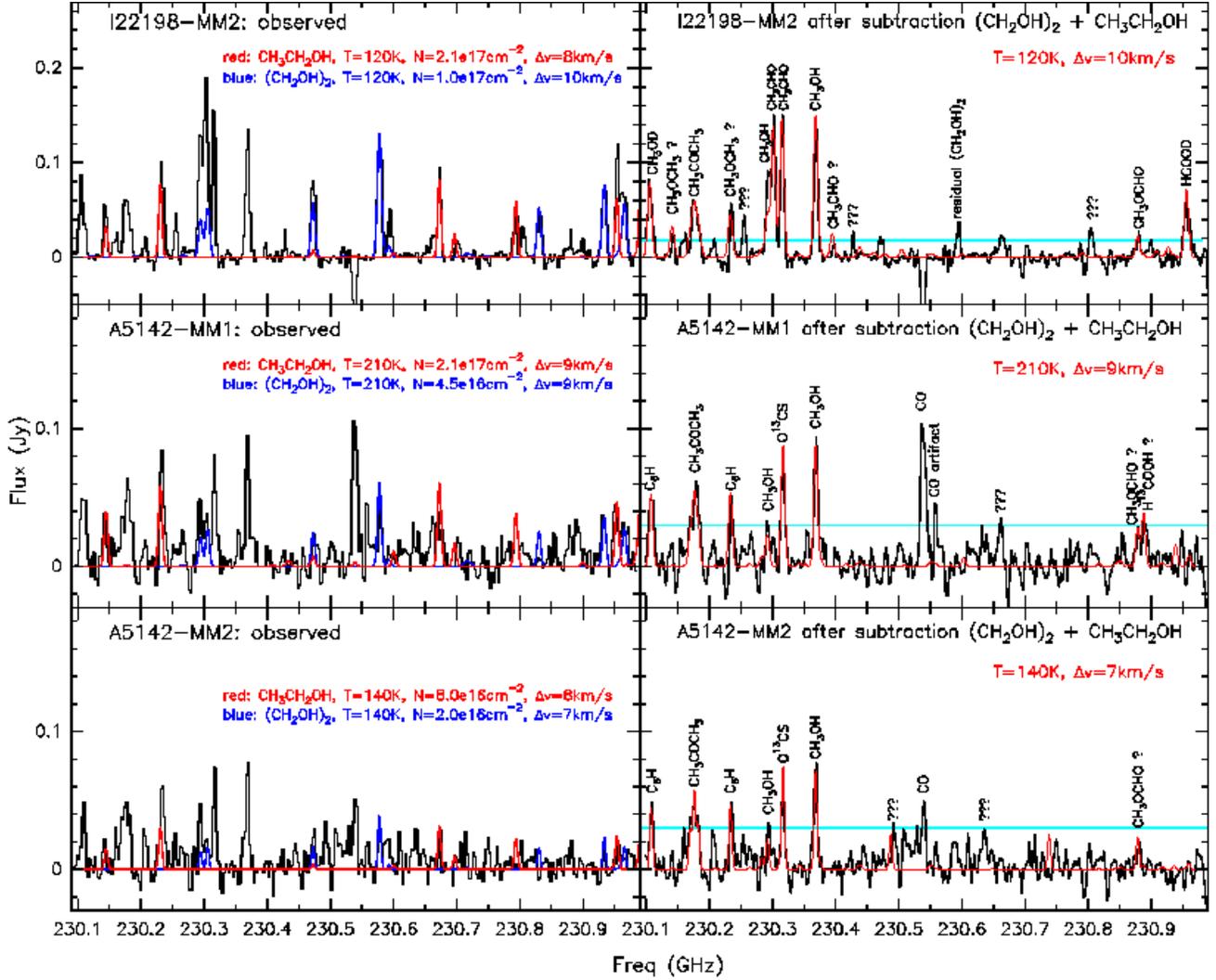, width=14cm, angle=270}\\
\end{tabular}
\caption{\emph{Left:} Spectrum for I22198 (top), A5142-MM1 (middle), and A5142-MM2 (bottom) over the PdBI continuum bands, with the computed synthetic spectra overlaid for \ethanol\ and \ethyleneg\ (see text). The spectrum is extracted by integrating the intensity over an area of $0.41''$, $0.52''$, $0.43''$ (FWHM), yielding conversion factors of 136.6, 84.9, and 124.2~K Jy$^{-1}$ for I22198, A5142-MM1, A5142-MM2, respectively.
                 \emph{Right:} Residual spectra after subtracting the contribution from \ethanol\ and \ethyleneg, with the identified lines marked at the top of each transition. The assumed temperatures and linewidths used to build the synthetic spectra are given in each panel, and the column densities used are listed in Table~1. The horizontal line indicates the 4$\sigma$ threshold used for line identification.
                 }
\label{flit3}
\end{center}
\end{figure*}

Among the strongest detected transitions we have chosen four to be representative of different excitation conditions, and computed the zero-order (integrated intensity) and first-order (velocity) maps (Figures~1-b--e and 2-b--e). From the figure it is seen that in I22198 the COM emission is restricted to MM2 without extending to MM2-S, and that this emission is elongated for almost all the molecules in the southeast-northwest direction (the only unresolved emission is that from \acetone, as in Orion-KL, \eg\ Friedel \et\ 2005). The deconvolved size and P.A. of the average emission of the four transitions shown in Fig.~1-b--e is listed in Table~1. 
It is interesting to note that the first-order moment map of the three resolved COM transitions of I22198 shows a velocity gradient in the direction perpendicular to outflow A (Fig.~1e).  Concerning A5142, COM emission is found in both millimeter continuum sources MM1 and MM2, with the first-order moment showing that there is a shift in velocities between the two sources of $\sim3$~\kms, as found by Zhang \et\ (2007), and with hints of an elongation in the east-west direction for MM1, and in the southeast-northwest direction for MM2 (Table~1). What is more,  the emission from MM1 reveals an extension to the southwest, apparent mainly in \methanol, which is following the CO redshifted emission shown in Fig.~2-e. 

\section{Discussion and Conclusions \label{sdis}}

\subsection{Hot corinos or massive hot cores?}



The set of detected molecules in the intermediate-mass hot cores of I22198 and A5142 does not include CN-bearing species. We inspected the observed frequency range and found that at least two transitions of \vinylcyanide\ should have been detected (at 230.488 and 230.739~GHz) with the same intensity, from synthetic spectra at rotational temperature in the range 100--600~K and column densities $\sim10^{15}$~cm$^{-2}$ (Table~1). Thus, the set of detected molecules is similar to the sets of the hot corino IRAS\,16293$-$2422, where \otcs, \acetaldehyde, and \methylformate\  are also detected (Bottinelli \et\ 2007; Caux \et\ 2011). However, IRAS\,16293$-$2422 is dominated by simple O-rich and HCO-rich species like H$_2$CO, SO$_2$, \methanol, 
\acetaldehyde, and \methylformate, while in I22198 and A5142 we detected more CH$_{2/3}$-rich molecules, such as \ethanol, \ethyleneg, and \acetone.
On the other hand,  massive hot cores,  such as Orion-KL (Caselli \et\ 1993; Blake \et\ 1996; Friedel \& Snyder 2008), G29.96$-$0.02 (Beuther \et\ 2007), or  G34.26+0.15 (Mookerjea \et\ 2007) show  strong emission in \vinylcyanide, and \ethylcyanide, as well as in \acetone\ and CH$_{2/3}$-rich
molecules (\eg\ \"Oberg \et\ 2011). Thus, the intermediate-mass hot cores in I22198 and A5142 show a chemistry with properties of both low-mass hot corinos (lack of CN-bearing molecules) and massive hot cores (CH$_{2/3}$-rich complex molecules). 

%

\begin{figure}[h]
\begin{center}
\begin{tabular}[b]{cc}
    \epsfig{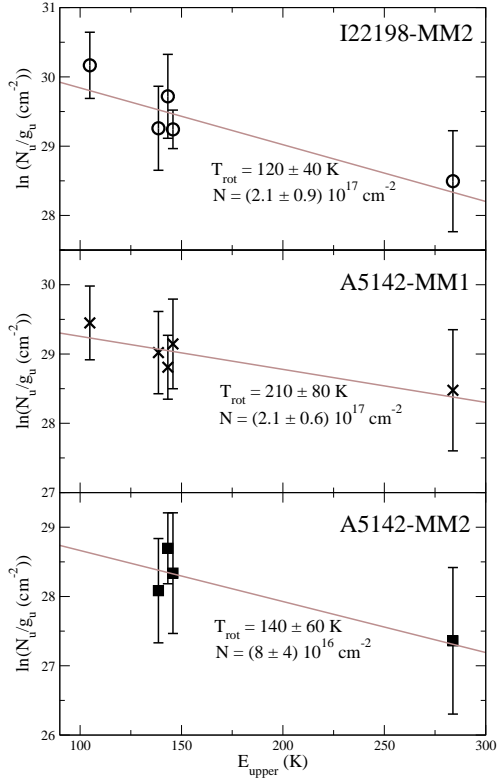}\\
\end{tabular}
\caption{Rotational diagram for \ethanol\ for I22198-MM2 (top), A5142-MM1 (middle), and A5142-MM2 (bottom).
For those \ethanol\ transitions which are significantly contaminated ($\sim50$\%) by emission from another molecule, half the line area is used (blended lines: 230.954~GHz (with energy of the upper state, $E_\mathrm{u}$, of 146~K) for I22198, and 230.231~GHz ($E_\mathrm{u}=143$~K) for A5142-MM1 and MM2.
}
\label{flit3}
\end{center}
\end{figure}

\subsection{Hot core emission at $\sim500$~AU: disks or outflows?}


The sizes (FWHM) of the COM emission range from 150--300~AU for I22198, and 300--1000~AU for A5142 (Table~1), and are, after Orion-KL (200--$1000$~AU: Guelin \et\ 2008, Favre \et\ 2011a, 2011b; Zapata \et\ 2011),  among the smallest in intermediate/high-mass star-forming regions ever measured ($>2000$~AU: \eg\ Beuther \et\ 2007; 
Liu \et\ 2010; 
Zapata \et\ 2010;
Beltr\'an \et\ 2011).
Thus, the present observations are well suited to consider whether the emission is due to internal heating (disks) and/or shocks (from outflows). 

\begin{table}[t!]
\caption{Properties of intermediate-mass hot cores at $\sim500$~AU  }
\centering
\begin{tabular}{l c c c}
\hline\hline\noalign{\smallskip}   
									&I22198-MM2			&A5142-MM1			&A5142-MM2		\\
\hline
\noalign{\smallskip}
distance (pc)							&760				&\multicolumn{2}{c}{1800}			\\
$L_\mathrm{bol}$ (\lo)					&370				&\multicolumn{2}{c}{2300}			\\
\hline
\multicolumn{3}{l}{\phn\phn \it continuum + outflow information}    \\
\hline
clustering (mm)							&no					&\multicolumn{2}{c}{yes}			\\
$L_\mathrm{cm}$ (mJy~kpc$^2$)			&0.34				&2.1					&$<0.9$			\\
outflow(s) PA ($^\circ$)					&$50$				&$-40,0$				&$45$	\\
$M_\mathrm{env}$ (\mo)$^\mathrm{a}$		&0.5--2				&2--9				&2--6			\\
$N_\mathrm{H_2}$ (cm$^{-2}$)$^\mathrm{a}$	&$2\times10^{25}$		&$1\times10^{25}$		&$2\times10^{25}$	\\
density (cm$^{-3}$)$^\mathrm{a}$			&$4\times10^9$\phe		&$1\times10^9$\phe		&$3\times10^9$\phe	\\
1.3 mm cont., RA						&22:21:26.753			&05:30:48.031			&05:30:48.024		\\
1.3 mm cont., DEC						&63:51:37.90\phn		&33:47:54.64\phn		&33:47:53.67\phn	\\
continuum size (AU)$^\mathrm{b}$			&$520\times300$		&$1200\times900$		&$1000\times400$		\\
continuum PA  ($^\circ$)$^\mathrm{b}$			&$-35$				&$-86$				&$+18$			\\
\hline
\multicolumn{3}{l}{\phn\phn \it molecular hot core (HC) information$^\mathrm{c}$}    \\
\hline
$T_\mathrm{rot, CH_3CH_2OH}$ (K)		&$\sim\!120$			&$\sim\!210$			&$\sim\!140$		\\
$N_\mathrm{CH_3CH_2OH}$ (cm$^{-2}$)	&$2.1\times10^{17}$	&$2.1\times10^{17}$	&$8.0\times10^{16}$	\\
$N_\mathrm{(CH_2OH)_2}$ (cm$^{-2}$)		&$1.0\times10^{17}$	&$4.5\times10^{16}$	&$2.0\times10^{16}$	\\
$N_\mathrm{CH_3COCH_3}$ (cm$^{-2}$)	&$5.0\times10^{16}$	&$1.0\times10^{17}$	&$4.0\times10^{16}$	\\
$N_\mathrm{CH_3OH}$ (cm$^{-2}$)		  	&$2.3\times10^{19}$	&$2.3\times10^{18}$	&$4.0\times10^{18}$	\\
$N_\mathrm{CH_3CHO}$ (cm$^{-2}$)		&$6.0\times10^{16}$	&$<1\times10^{16}$		&$<7\times10^{15}$	\\
$N_\mathrm{HCOOD}$ (cm$^{-2}$)			&$3.0\times10^{16}$		&$<2\times10^{16}$	&$<1\times10^{16}$	\\
$N_\mathrm{CH_3OCHO}$ (cm$^{-2}$)		&$2.5\times10^{17}$		&$<2\times10^{17}$		&$<2\times10^{17}$	\\
$N_\mathrm{C_6H}$ (cm$^{-2}$)			&$<6\times10^{15}$	&$3.0\times10^{15}$		&$5.0\times10^{15}$	\\
$N_\mathrm{O^{13}CS}$ (cm$^{-2}$)		&$<5\times10^{17}$	&$1.5\times10^{16}$		&$1.0\times10^{16}$	\\
$N_\mathrm{CH_2CHCN}$ (cm$^{-2}$)		&$<2\times10^{15}$		&$<3\times10^{15}$		&$<3\times10^{15}$	\\
HC offset$^\mathrm{d}$					&$-0\farcs08$,$+0\farcs01$&$-0\farcs27$,$-0\farcs20$	&$-0\farcs12$,$-0\farcs21$\\
HC size (AU)$^\mathrm{b}$				&$280\times160$\phn	&$1000\times<380$		&$820\times340$		\\
HC PA ($^\circ$)$^\mathrm{b}$				&$+155$				&$+89$				&$+120$			\\
\hline
\end{tabular}
\begin{list}{}{}
\item[$^\mathrm{a}$] Masses computed assuming a dust temperature in the range 30--100~K, and the dust opacity law given in the main text. The H$_2$ column density and density, estimated assuming spherical symmetry and uniform density, are calculated adopting a dust temperature of 50~K, and the mass and size from the continuum emission.
\item[$^\mathrm{b}$] Deconvolved size and P. A. For the case of the COMs, the (average) size is obtained from Gaussian fits to the zero-order moment map integrated in the range 230.1--230.4~GHz.
\item[$^\mathrm{c}$] Column densities are calculated assuming the rotational temperature and linewidth given in Fig.~4, and are corrected for filling factor (source sizes from Fig.~3).
\item[$^\mathrm{d}$] RA and DEC offset with respect to continuum peak position.
\end{list}
\label{timhcs}
\end{table}



For the case of I22198, the continuum and COM emission peaks are coincident within $<0.1''$ (Table~1), and COM emission is not present near MM2-S, which in turn is coincident with an IRAC source at 3.6~\mum\ and 4.5~\mum. This is suggestive of MM2 being less evolved than MM2-S. On the other hand, the two outflows present in the region show different properties, 
with HCN and SiO emission detected only along the direction of outflow A
 (Fig.~1a, S\'anchez-Monge 2011), which supports the interpretation of two different outflows against one single wide-angle outflow. Thus, a plausible interpretation is that MM2 is the driving source of outflow A. If this is the case, the COM emission, with a size of 300~AU, would be tracing a rotating disk, similar to the disk toward the NGC\,1333-IRAS2A hot corino, of 200~AU of size (Joergensen \et\ 2005). 
The velocity gradient seen in the COM emission of MM2 yields 
a central protostellar mass of $\gtrsim4.2$~\mo\ (following S\'anchez-Monge \et\ 2010; lower limit from assuming an inclination angle with respect to the plane of the sky of $90^\circ$), supporting the intermediate-mass nature of the object.
From a theoretical point of view, models of disks in massive protostars predict disk sizes of hundreds of AU, and linewidths around 10~\kms\ (\eg\ Krumholz \et\ 2007), both parameters coincident with our measurements. 
Thus, although the COM emission could be contaminated by shocks from the outflows,
our observations in I22198 are consistent with the COM emission arising from a disk. 


Concerning A5142,  the COM emission is offset from the continuum emission by $\sim0.3''$ (or $\sim500$ AU, Table~1) for both MM1 and MM2. 
Possible explanations are: i) COM emission is optically thick; ii) COM emission is strongly affected by shocks and the passage of outflow(s);  iii) the continuum and COM peaks are tracing different objects.
We discard the first possibility because the peak of COMs for which a rotational diagram was performed is offset from the continuum peak and the rotational diagram does not show a clear sign of high opacity (Fig.~4). 
%
The second possibility of COM emission produced by shocks/outflows is questioned by the fact that the peak of COMs near A5142-MM1 is coincident with the peak of a centimeter source (Fig.~2-d), and the centimeter source, most likely tracing a thermal radiojet, falls exactly at the center of symmetry of outflow C, as traced by \water\ masers (see Figure 3 in Goddi \& Moscadelli 2006). This suggests that the COM emission near MM1 could be associated with a possible disk (perpendicular to the centimeter source elongation) from the driving source of outflow C,
which is consistent with recent \methanol\ and \water\ maser observations
(Goddi \et\ 2011b).
In addition, the CO\,(2--1) emission of outflow B seems to have its center of symmetry in MM2 (Fig.~2-a), as already suggested by Goddi \& Moscadelli (2006), and interestingly the COM emission near MM2 is elongated in the direction perpendicular to outflow B (Fig.~2-b,d; Table~1), suggesting again the association with a disk. 
%

To further assess the nature of the COM emission in A5142-MM1 and A5142-MM2 we have computed the ratio of a HCO-rich species with respect to a CH$_{2/3}$-rich species, expected to be small for processed (disk) material (\eg\ \"Oberg \et\ 2011).  The \acetaldehyde/\ethanol\ ratio is $<0.05$ for A5142-MM1, and $<0.09$ for A5142-MM2, about one order of magnitude smaller than those derived in shocks associated with outflows, and similar to the values derived in hot corinos and hot cores (\eg\ Arce \et\ 2008; \"Oberg \et\ 2011).
This gives further support to the interpretation of COM emission in A5142-MM1 and MM2 being associated with disks, although obviously some (minor) contribution from shocks/outflows could still be present. 
%
Finally, the elongation of A5142-MM1 to the southwest could be arising in shocks/outflows, given its strong similarity to the CO\,(2--1) redshifted lobe (Fig.~2-e), and the fact that its emission is detected in \otcs\  and \methanol, which are molecules typically found in chemically rich outflows (\eg\ Bachiller \& P\'erez-Guti\'errez 1997; Leurini \et\ 2011).

In conclusion, 
the high angular resolution PdBI observations have revealed that the hot cores associated with the intermediate-mass protostars I22198 and A5142 are most likely tracing disks, with some contribution from the shocked gas of the outflows.

\acknowledgments
\begin{small}
A.P. is grateful to  Catherine Walsh and Bel\'en Tercero for useful discussions on modeling of protostellar disks and line identification, and to the anonymous referee, whose comments largely improved the quality of this letter.  A.P. is  supported by the Spanish MICINN grant AYA2008-06189-C03 (co-funded with FEDER funds) and by a JAE-Doc CSIC fellowship co-funded with the European Social Fund. This paper was partially supported by the Spanish MICINN program ``CONSOLIDER INGENIO 2010: Molecular Astrophysics,
Herschel-ALMA Era, ASTROMOL'' (CSD2009-00038), and by the European Community's Seventh Framework Program (FP7/2007--2013) under agreement 229517. 
\end{small}



\end{document}